\documentclass[12pt,a4paper]{article}

\usepackage[dvips]{graphicx} % Поддержка вставки графических файлов.

\newcommand\ba{\begin{eqnarray}}
\newcommand\ea{\end{eqnarray}}
\newcommand\baf{\begin{eqnarray}}   % Уравнение с рамочкой.
\newcommand\eaf{\end{eqnarray}}
\begin{document}
\textwidth=135mm
 \textheight=200mm

\begin{center}
{\bfseries Hadron structure and spin effects in elastic hadron scattering at NICA energies
%Spin effects in elastic scattering at NICA energies
\footnote{{\small e-mail: selugin@theor.jinr.ru
}}}
\vskip 5mm
O.V. Selyugin
\vskip 5mm
{\small {\it  Joint Institute for
Nuclear Research, 141980 Dubna, Russia}} \\
\end{center}
\vskip 5mm
\centerline{\bf Abstract}
 The spin effects in the elastic proton-proton scattering are analysed at NICA energies.
  It is shown the importance the investigation of the region of the diffraction minimum
  in the differential cross sections. Some possible estimation of spin effects are given for
  the different NICA energies in the framework of the new high energy generelazed structure
  (HEGS) model.
 % \vskip 10mm
  \vskip 4mm
  PACS:
      {13.40.Gp}, %{discribing text of that key} \and
      {14.20.Dh}, %{text} % \and
      {12.38.Lg}
  \vskip 6mm

%\label{sec:intro}
\section*{Introduction}
\label{intro}
  One of the most important tasks of modern physics is the research into the basic properties of
  hadron interaction.
	The dynamics of strong interactions  finds its most
  complete representation in elastic scattering. % at small angles.
  It is just this process that allows the verification of the
results obtained from the main principles of quantum field theory:
the concept of the scattering amplitude as a unified analytic function
of its kinematic variables connecting different reaction channels
were introduced in the dispersion theory by N.N. Bogoliubov\cite{bog}.
 Now many questions of  hadron interactions are connected with
 modern problems of  astrophysics such as unitarity and the optical theorem \cite{COTH-20},
 and problems of baryon-antibaryon symmetry and CP-invariance violation \cite{Uzik}
	%  The dynamics of strong interactions  finds its most
 % complete representation in elastic scattering at small angles.
 The main domain of  elastic scattering is  small angles.
  Only in this region of interactions can we measure the basic properties that
  define the hadron structure. %: the total cross section,
  % the slope of the diffraction peak and the parameter $\rho(s,t) $.
  Their values
  are connected, on the one hand, with the large-scale structure of hadrons and,
  on the other hand, with the first principles that lead to the
  theorems on the behavior of scattering amplitudes at asymptotic
  energies \cite{mart,roy}.

  Modern studies
   of elastic scattering of high energy protons lead to several unexpected results reviewed, e.g., in \cite{Drem1,Drem2}.
    Spin amplitudes of the elastic $NN$ scattering constitute a spin picture % portrait
     of the nucleon.
   Without knowledge of the spin $NN$-amplitudes  it is not possible to understand spin observable
   of  nucleon scattering off nuclei.
       In the modern picture, the structure of hadrons is determined by Generalized Distribution functions (GPDs),
       which include the corresponding parton distributions (PDFs). The sum rule \cite{Ji97}
       allow to obtain the elastic form factor (electromagnetic and gravitomagnetic)
       through the first and second integration moments of GPDs. It leads to remarkable properties of GPDs -
       some corresponding to inelastic  and elastic scattering of hadrons.
 Now some different
models examining the nonperturbative instanton contribution lead
to sufficiently large spin effects at superhigh energies \cite{Anselmino,Dor-Koch} %  {Forte},
%\cite{Dor}.
The research of such spin effects
will be a crucial stone for different models and will help us
to understand the interaction and structure of particles, especially at large
distances.
There are large programs of researching spin effects
at different  accelerators. Especially, we should like to note
the programs at  NICA, %RHIC \ci{rhic},
where the polarization of both the collider beams will be constructed.
So it is very important to obtain reliable predictions for the spin
asymmetries at these energies. In this paper, we extend the model
predictions to spin asymmetries in the NICA energy domain.

   The NICA SPD detector  bounded a very small momentum transfer.
   If in the first steps the angles start from $16$ mrad, then the minimum
   momentum transfer, which can be measured is more than $-0.01$ GeV$^2$. Hence it is needed to exclude the
   Coulomb-nuclear interference region, where the real part of the spin-non-flip amplitude can be determined,
%   end the size of the $A_N$ can be determined.
  We should  move our researches on
   the region of the diffraction minimum, where the imaginary part of the spin-non-flip amplitude
   changes its sign. Note that in some models the absence of the second diffraction minimum
   is explained by the contribution in the differential cross section of the spin-flip amplitude \cite{ETT-79}
   The interference of the hadronic
 and electromagnetic amplitudes may give an important contribution not
 only at very small transfer momenta but also in the range of the
 diffraction minimum \cite{Selphase}.  However, for that one should know the
 phase of the interference of the Columbic and hadronic amplitude at
 sufficiently large transfer momenta too.

   Using the existing model of nucleon elastic scattering
   at high energies $\sqrt{s}>9$ GeV - 14 TeV \cite{HEGS0,HEGS1}, which involves
   minimum of free parameters, we are going to develop its extended version aimed  to describe all available
   data on cross sections and spin-correlation parameters  at  lower energies down  to the  SPD NICA region.
   The model will be based on the usage  of known information on generalized
    parton distributions in the nucleon, electro-magnetic and gravitomagnetic  form factors of the nucleon
    taking into account  analyticity and unitarity requirements and providing compatibility with the
    high energy limit,  where the pomeron exchange dominates.

\section{%The elastic nucleon scattering in the framework of the HEGS model in the dip region
          HEGS model and spin effects in the dip region of momentum transfer}
%\section{Model approximation}

  The differential cross sections of nucleon-nucleon elastic scattering  can be written as a sum of different
  helicity  amplitudes:
%\begin{eqnarray}
%  \frac{d\sigma}{dt} =
% \frac{2 \pi}{s^{2}} (|\phi_{1}|^{2} +|\phi_{2}|^{2} +|\phi_{3}|^{2}
%  +|\phi_{4}|^{2}
%  +4 | \phi_{5}|^{2} ). \label{dsdt}
%\end{eqnarray}
% The differential cross section is
 \begin{eqnarray}
   \frac{d \sigma}{dt} = \frac{2 \pi}{s^2} ( |\Phi_{1}|^2 + |\Phi_{2}|^2+
   |\Phi_{3}|^2 + |\Phi_{4}|^2+4 |\Phi_{5}|^2.
\end{eqnarray}

%\ba
%  A_N\frac{d\sigma}{dt}&=& -4\pi Im[(F_1+F_2+F_3-F_4)*F_5^{*}).  \lab{an}
%\ea
%\ba
%  A_N\frac{d\sigma}{dt}&=& -4\pi Im[(F_1+F_2+F_3-F_4)*F_5^{*}).  \lab{an}
%\ea
\begin{eqnarray}
 A_{N} \frac{d \sigma}{dt} = -\frac{4 \pi}{s^2}
   [ Im (\Phi_{1}(s,t) + \Phi_{2}(s,t)+ \Phi_{3}(s,t) - \Phi_{4})(s,t) \Phi^{*}_{5}(s,t)]
\end{eqnarray}
and
\begin{eqnarray}
  A_{NN} \frac{d \sigma}{dt} = \frac{4 \pi}{s^2}
   [ Re (\Phi_{1}(s,t) \Phi^{*}_{2}(s,t) - \Phi_{3}(s,t) \Phi^{*}_{4})(s,t) + |\Phi_{5}(s,t)|^2]
\end{eqnarray}

  The HEGS model \cite{HEGS0,HEGS1} takes into account all five spiral electromagnetic amplitudes.
   The electromagnetic amplitude can be calculated in the framework of QED.
    In the high energy approximation, it can be  obtained \cite{bgl}
  for the spin-non-flip amplitudes:
  \begin{eqnarray}
  F^{em}_{1}(t) = \alpha f_{1}^{2}(t) \frac{s-2 m^2}{t}; \ \ \  F^{em}_3(t) = F^{em}_1;
%   \\ \nonumber
  \end{eqnarray}
   and for the spin-flip amplitudes:
   with the  electromagnetic and hadronic interactions included, every amplitude $\phi_{i}(s,t)$
  can be described as
\begin{eqnarray}
  \phi_{i}(s,t) =
  F^{em}_{i} \exp{(i \alpha \varphi (s,t))} + F^{h}_{i}(s,t) ,
\end{eqnarray}
where
%\begin{eqnarray}
 $  \varphi(s,t) =  \varphi_{C}(t) - \varphi_{Ch}(s,t)$, and
%\end{eqnarray}
   $ \varphi_{C}(t) $ will be calculated in the second Born approximation
 in order  to allow the evaluation   of   the Coulomb-hadron interference term $\varphi_{Ch}(s,t)$.
   The  quantity $\varphi(s,t)$
 has been calculated at large momentum transfer including
  the region of the diffaction minimum %and discussed  by many authors (see
  \cite{selmp1,selmp2,Selphase} and references therein.

 Let us  define the hadronic spin-non-flip amplitudes as
\begin{eqnarray}
  F^{h}_{\rm nf}(s,t)
   &=& \left[\Phi_{1}(s,t) + \Phi_{3}(s,t)\right]/2; \label{non-flip}
 \end{eqnarray}
  The model is based on the idea that at high energies a hadron interaction in the non-perturbative regime
      is determined by the reggenized-gluon exchange. The cross-even part of this amplitude can have two non-perturbative parts, possible standard pomeron - $(P_{2np})$ and cross-even part of the 3-non-perturbative gluons ($P_{3np}$).
      The interaction of these two objects is proportional to two different form factors of the hadron.
      This is the main assumption of the model.
      The second important assumption is that we choose the slope of the second term
       four times smaller than the slope of the first term, by  analogy with the two pomeron cuts.
      Both terms have the same intercept.

      The form factors are determined by the Generalized parton distributions of the hadron (GPDs).
      The first form factor corresponding to the first momentum of GPDs is the standard electromagnetic
      form factor - $G(t)$. The second form factor is determined by the second momentum of GPDs -$A(t)$.
      The parameters and $t$-dependence of  GPDs are determined by the standard parton distribution
      functions, so by  experimental data on  deep inelastic scattering and by  experimental data
      for the electromagnetic form factors (see \cite{GPD-ST-PRD09}). The calculations of the form factors were
      carried out in \cite{GPD-PRD14}.
The final elastic  hadron scattering amplitude is obtained after unitarization of the  Born term.
     At large $t$  our model calculations are extended up to $-t=15 $ GeV$^2$.
  We added a small contribution of the energy  independent  part
  of the spin flip amplitude in the form similar to the proposed    in \cite{G-Kuraev2}
  and analyzed in \cite{HEGS-SP}.
  \begin{eqnarray}%\ba
  F_{sf}(s,t) \ =  h_{sf} q^3 F_{1}^{2}(t) e^{-B_{sf} q^{2}}.
% \nonumber
  \end{eqnarray}
    The energy dependent part of the spin-flip amplitude is related to the main amplitude
    but with an additional kinematic factor and the main slope taken twice more,
    conformity with the paper \cite{PS-EPA02,cudpr-epja}.     The form factors incoming in the
    spin-flip amplitude are determined by the GPD functions $H(s,t,x)$ and    $E(s,t,x)$,
    which include the corresponding PDF distributions.
  The model is very simple from the viewpoint of the number of fitting parameters and functions.
  There are no any artificial functions or any cuts which bound the separate
  parts of the amplitude by some region of momentum transfer.

 Now we shell restrict our discussion to the analysis of  $A_N$ as there are some experimental data
 in the region of NICA energies.
   In the standard pictures the spin-flip and double spin-flip amplitudes
    correspond to the spin-orbit $(LS)$ and spin-spin $(SS)$ coupling terms.
  The contribution to
  $A_N$ from the hadron double spin-flip amplitudes
   already at $p_L = 6 \ $GeV/c is of the second order
  compared to the contribution from the spin-flip amplitude.
   So with the usual high energy approximation for the helicity amplitudes
   at  small transfer momenta we suppose that
   $\Phi_{1}=\Phi_{3}$ and we can neglect the contributions of the hadron parts
   of $\Phi_2-\Phi_4$.
   Note that if $\Phi_{1}, \Phi_3, \Phi_5$ have the same phases, their interference contribution
   to $A_N$ will be zero, though the size of the hadron spin-flip amplitude can be large.
   Hence, if this phase has  different $s$ and $t$ dependencies, the contribution from the hadron
   spin-flip amplitude to $A_N$ can be zero at $s_i, \ t_i$ and non-zero at other $s_j, \ t_j$.
e experimental data ($\sum \chi^2/n_{dof} =1.24$).

       %==================Figs.1 =============================
%\label{sec:figures}
%~\vspace{-1.5cm}
\begin{figure*}
\begin{center}
\includegraphics[width=0.4\textwidth] {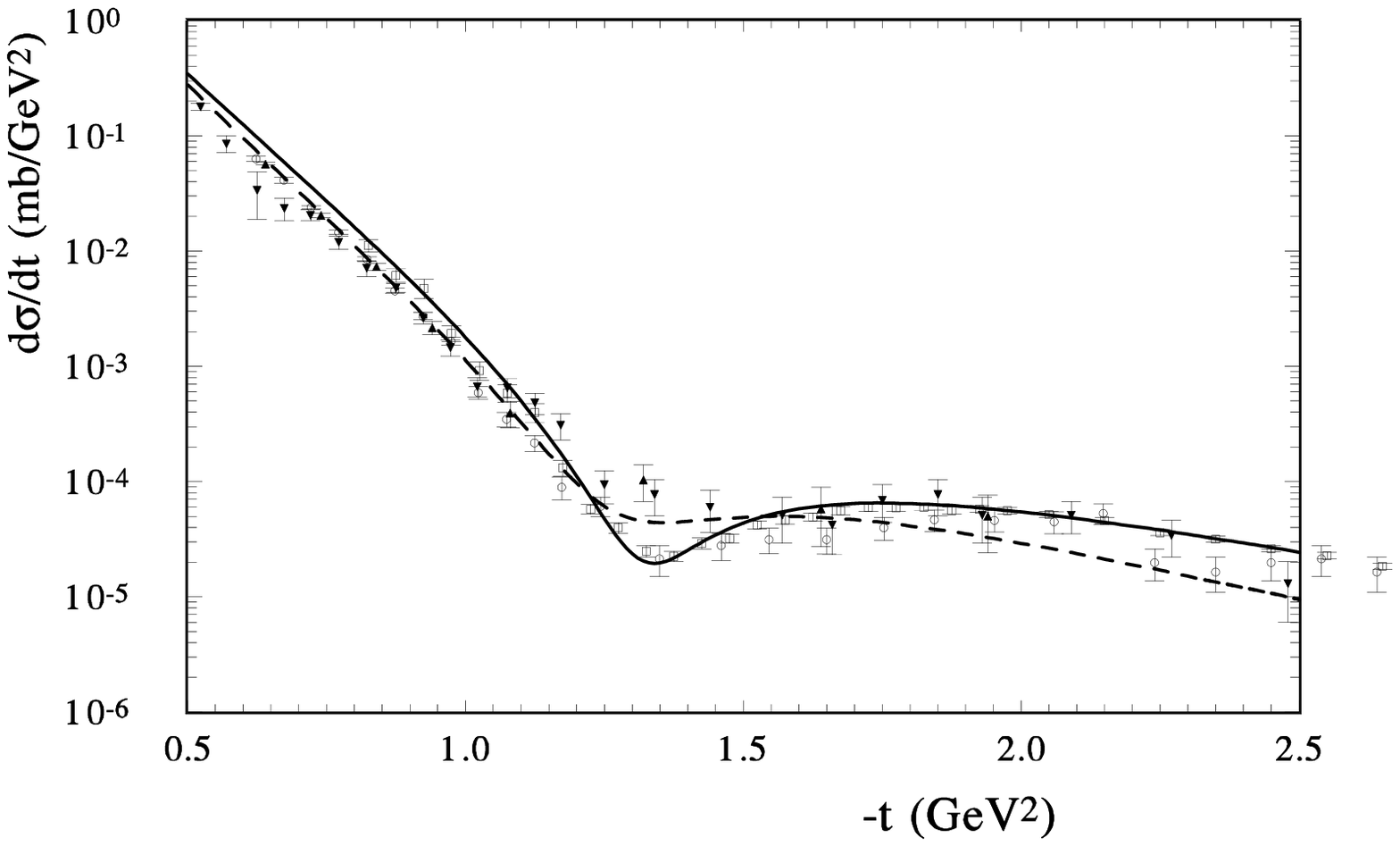}
\includegraphics[width=0.4\textwidth] {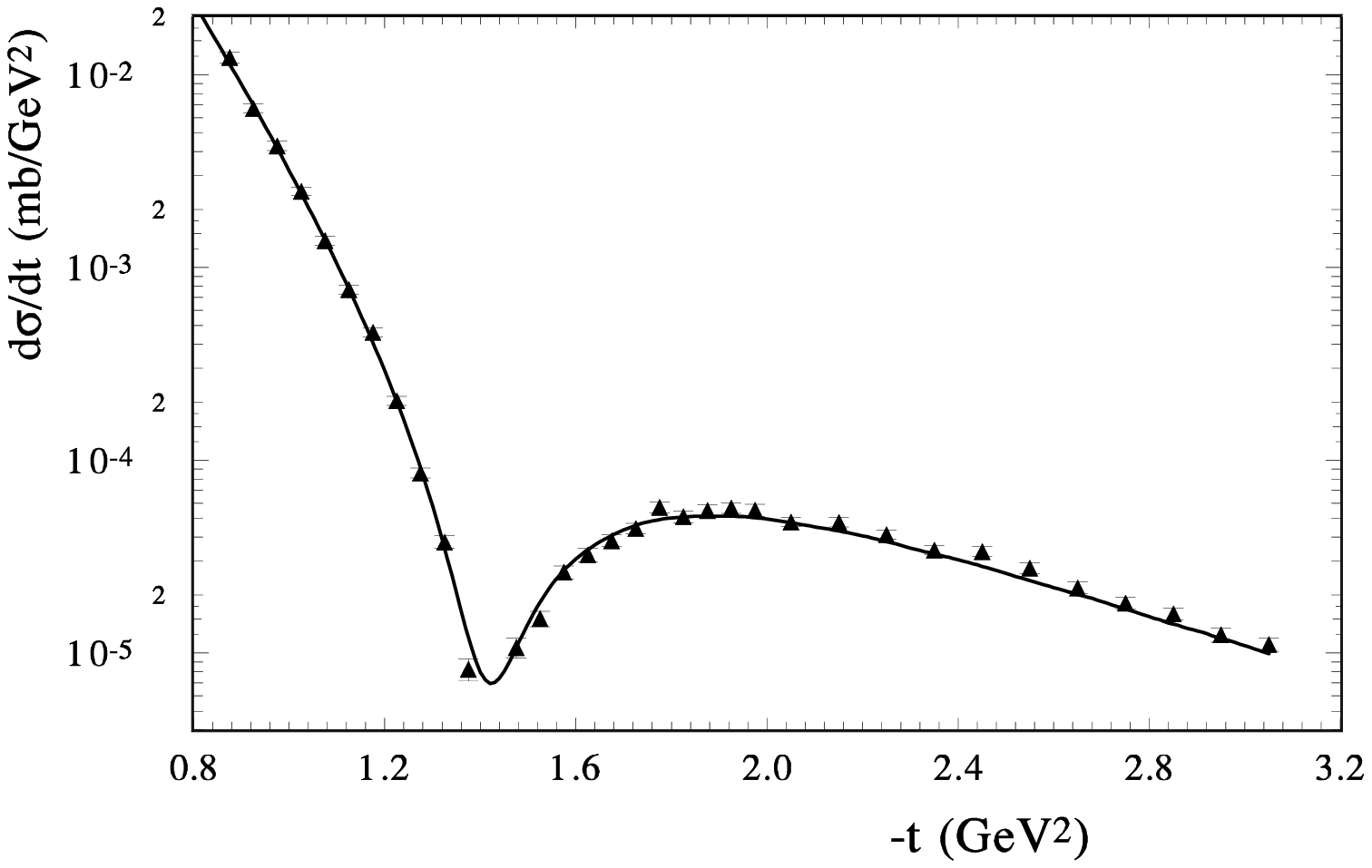}
\end{center}
%\includegraphics[width=0.5\textwidth] {asp52p8sl.ps}
%\includegraphics[width=0.5\textwidth] {as52s3b.ps}
%\vspace{0.5cm}
\caption{ The model calculation of the diffraction minimum in $d\sigma/dt$ of  $pp $ scattering
   [left] at  $  \sqrt{s}=30.4  $~GeV;
 [right] for $pp$ and $p\bar{p}$  at $\sqrt{s}=52.8$ GeV \cite{data} % proton-proton and proton-antiproton
  scattering.
  % (sqwords, circle- experimental data $pp$ and triangles $p\bar{p}$  correspondingly )% \cite{data}); (lines - the model calculations).
 }
% (left) and for $p\bar{p}$ (right). }\label{Fig:MV}
\end{figure*}
%

%==================Figs.2 =============================
%\label{sec:figures}
%~\vspace{-1.5cm}
\begin{figure*}
\begin{center}
\vspace{-1.5cm}
\includegraphics[width=0.6\textwidth] {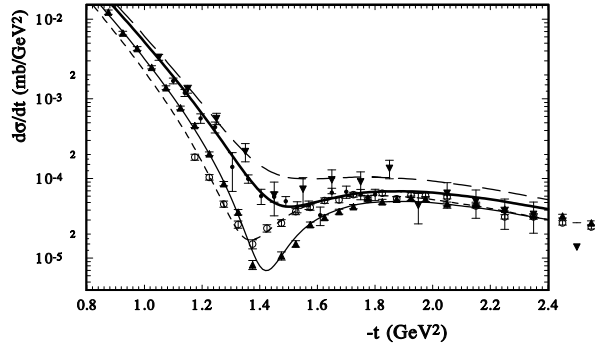}
\end{center}
\vspace{1.cm}
\caption{ The model calculation of the diffraction minimum in $d\sigma/dt$ of  $pp $
   at  $  \sqrt{s}=13.4; 16.8;  30.4; 44.7;  $~GeV;
 (lines, respectively,  long dash;  solid; thin-solid, and short - dash ); %solid+ants);
and experimental data \cite{data}, respectively,  - the triangle down,  the circles (solid),
  triangle up, and  circles    ) %  \cite{T7a}.
% , normalized on the theoretical curve.
% [right] at $\sqrt{s}=52.8$ GeV for proton-proton and proton-antiproton (squares, circle- experimental data $pp$ and triangles $p\bar{p}$  correspondingly )% \cite{data})
 }
% (left) and for $p\bar{p}$ (right). }\label{Fig:MV}
\end{figure*}

 Now let us examine the form of the differential cross section in the region of the momentum transfer
     where the diffractive properties of
     elastic scattering appear most strongly - it is the region of the diffraction dip.
 The form and the energy dependence of the diffraction minimum  are very sensitive
   to different parts of the scattering amplitude. The change of the sign of the imaginary part
   of the scattering amplitude determines the position of the minimum and its movement
    with changing  energy.
   The contributions of the real part of the spin-non-flip scattering amplitude and
   the square of the spin-flip amplitude  determine the size and the energy dependence of the dip.
   Hence, this depends    heavily on the odderon contribution.
   The spin-flip amplitude gives the contribution to the differential cross  sections additively.
   So the measurement of the form and energy dependence of the diffraction minimum
   with high precision is an important task for future experiments.

The HEGS model reproduces   $d\sigma/dt$ at very small and large $t$ and provides a qualitative description of the dip region
at $-t \approx 1.6 $~GeV$^2 $, for $\sqrt{s}=10 $~GeV and  $-t \approx 0.45 $~GeV$^2 $   for $\sqrt{s}=13 $~TeV. %(Fig.2).
 Note that it gives a good description for the proton-proton and proton-antiproton elastic scattering
  or $\sqrt{s}=53 $~GeV and for $\sqrt{s}=62.1 $~GeV (Fig.2a).

  The dependence of the position of the diffraction minimum on $t$ is determined in most part by the growth of the total cross sections  and the slope of the imaginary part of the scattering amplitude.
  Figures 2b and 3  show this a dependence obtained in the HEGS model at different energies.
%   in   the huge energy interval.
 %  The energy dependence of  the position of the diffraction minimum $t_{min}(s,t)$
 %  can be reproduced by a simple approximation, with a right value at $s \rightarrow \infty$,

%\vspace{-2cm}
\begin{figure*}
\begin{center}
\includegraphics[width=0.4\textwidth] {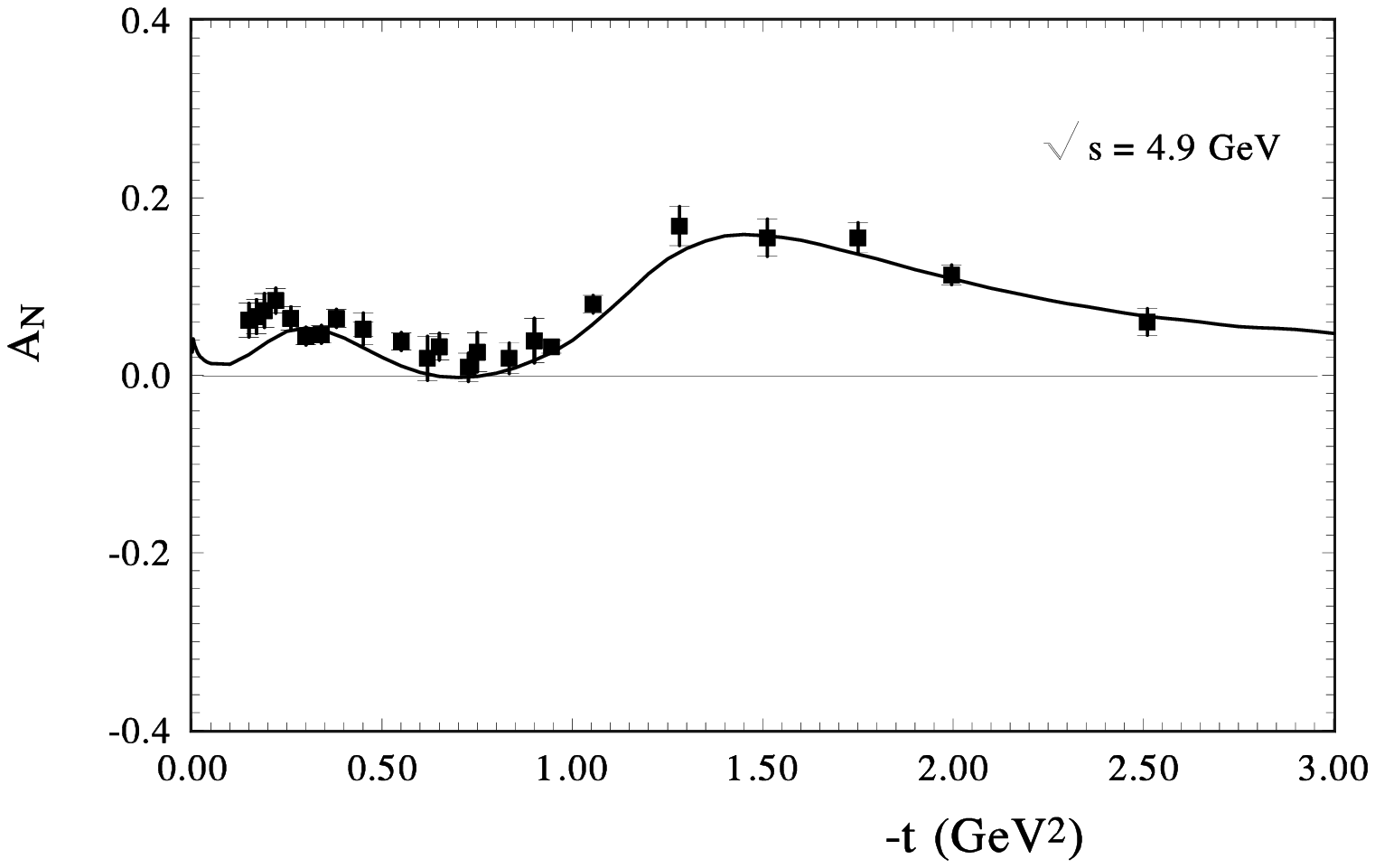}
\includegraphics[width=0.4\textwidth] {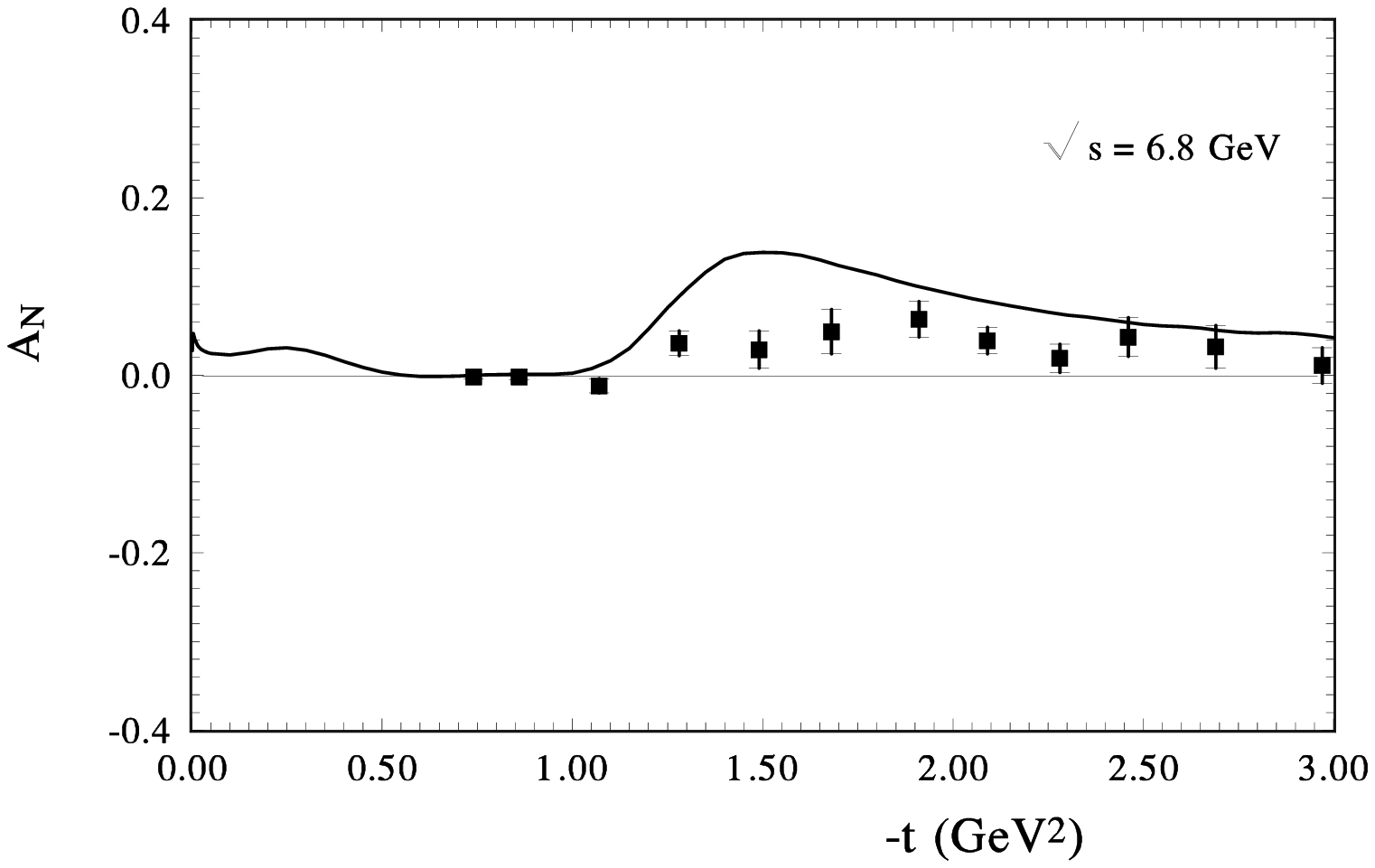}
\end{center}
\caption{The analyzing power $A_N$ of pp - scattering
      calculated:
  a) at $\sqrt{s} = 4.9 \ $GeV  (the experimental data \cite{Pol4p9}),
   and
  b) at $\sqrt{s} = 6.8 \ $GeV    (points - the existence experimental data \cite{Pol6p8} ).
  %    (the experimental data \cite{plb,pl12})
   }
\label{fig:1}       % Give a unique label
\end{figure*}

\begin{figure*}
\begin{center}
\includegraphics[width=0.4\textwidth] {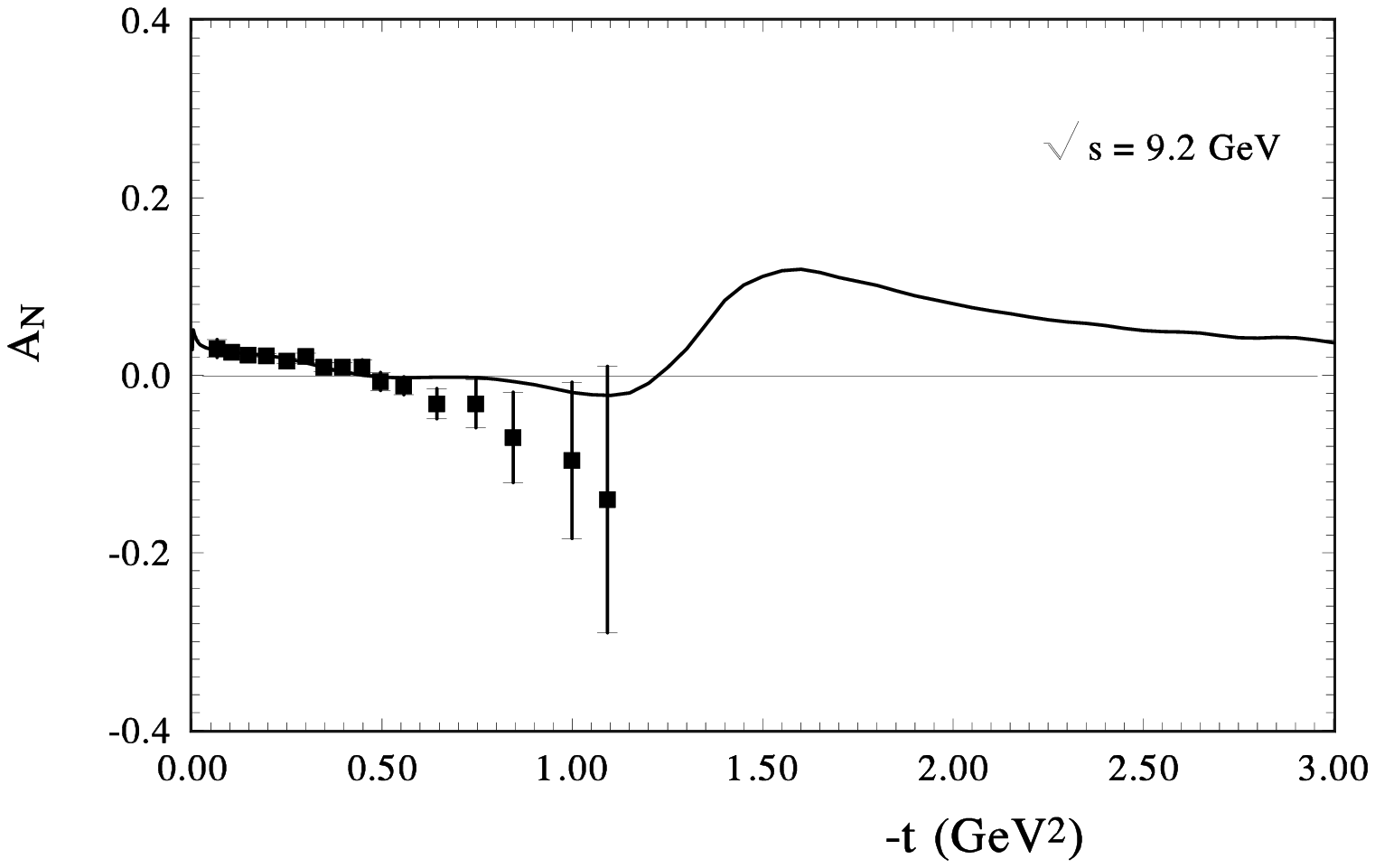}
\includegraphics[width=0.4\textwidth] {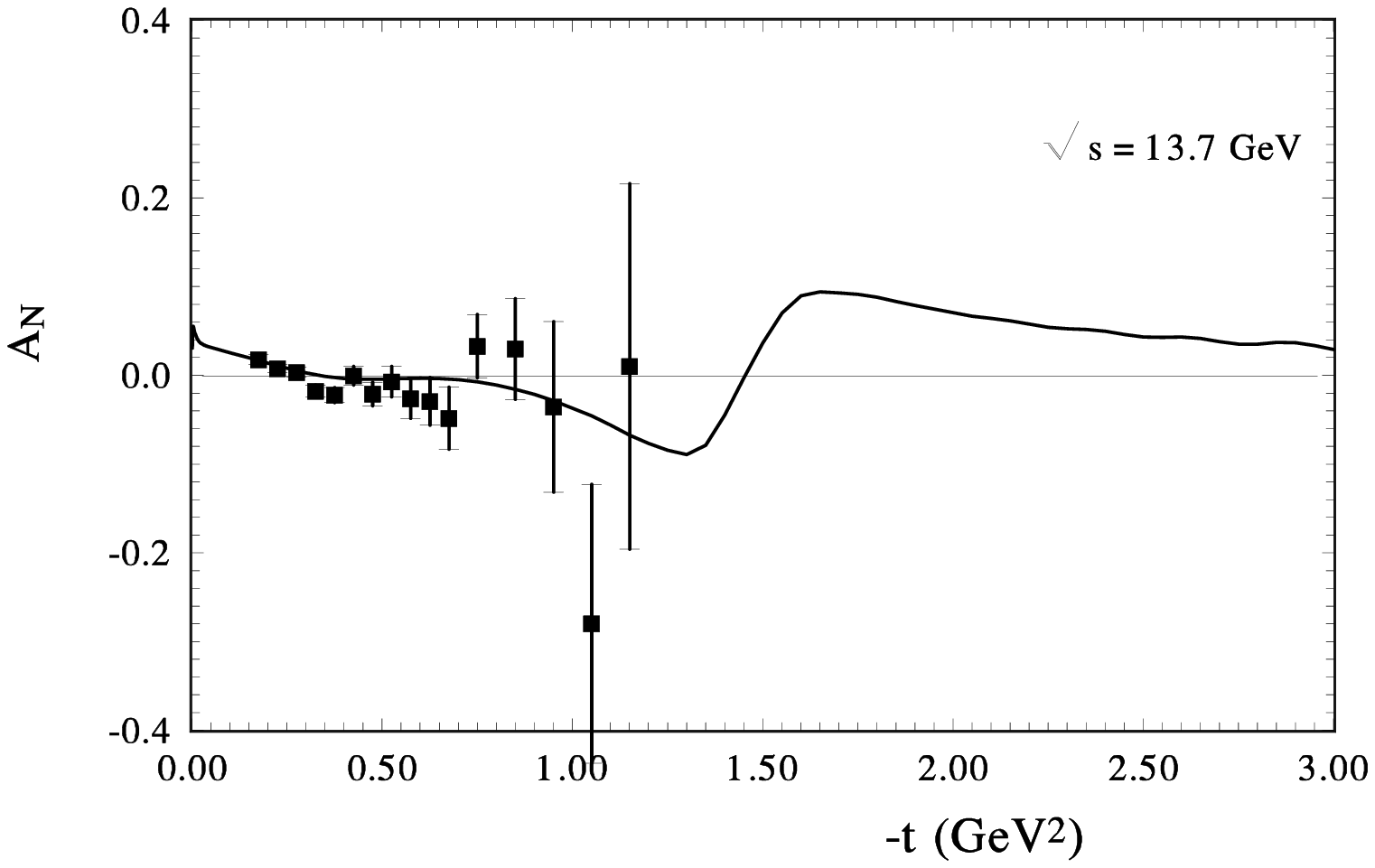}
\end{center}
\caption{The analyzing power $A_N$ of pp - scattering
      calculated:
   a) at $\sqrt{s} = 9.2 \ $GeV, and  (the experimental data \cite{Pol9p2}), and
   b)    at $\sqrt{s}= 13.7 \ $GeV (points - % the existence experimental data).
       the experimental data \cite{Pol23p4}).
          }
\label{fig:1}       % Give a unique label
\end{figure*}

     In Fig.3, the description of the diffraction minimum in our model is shown for NICA energies.
     The HEGS model reproduces  sufficiently well  the energy dependence and the form of the diffraction dip.
     In this energy region the diffraction minimum reaches the sharpest dip at  $\sqrt{s}=30 $~GeV  near
     the final NICA energy.
     Note that at this energy the value of $\rho(s,t=0)$ also changes its sign in the proton-proton
     scattering.

The calculated analyzing power  at $p_L = 6 \ GeV/c$ is shown in
  Fig.4a.
  One can see that a good description %, practically the same as in \cite{wak},
   of experimental data on the analyzing power
  can be reached only with
   one  hadron-spin flip amplitude.

    The experimental data at $p_L = 11.75 \ $GeV/c  seriously
 differ from those at $p_L = 6 \ $GeV/c but our calculations
 reproduce $A_{N}$ sufficiently well (Fig.4b ).
 It is shown that our energy dependence of the spin-flip amplitudes
  was chosen correctly and we may hope that  further we will obtain
  correct values of the analyzing power and other spin correlation parameters.

  From Fig.4 we can see that in the region
   $|t| \approx  0.2  \div 1 \ $ GeV$^2$ the contributions from the hadron spin-flip amplitudes
  are most important.
  At last, Fig.5a shows our calculations at $p_L = 200 \
  GeV/c$.

  At this energy, the contributions of the phenomenological energy independent part
  of the spin-flip amplitude is  compared with the energy dependent part.
  The spin effect is sufficiently large and has a specifical form,
   which is determined by the form of the differential
  cross section in the diffraction dip domain.

\begin{figure*}
\begin{center}
\includegraphics[width=0.4\textwidth] {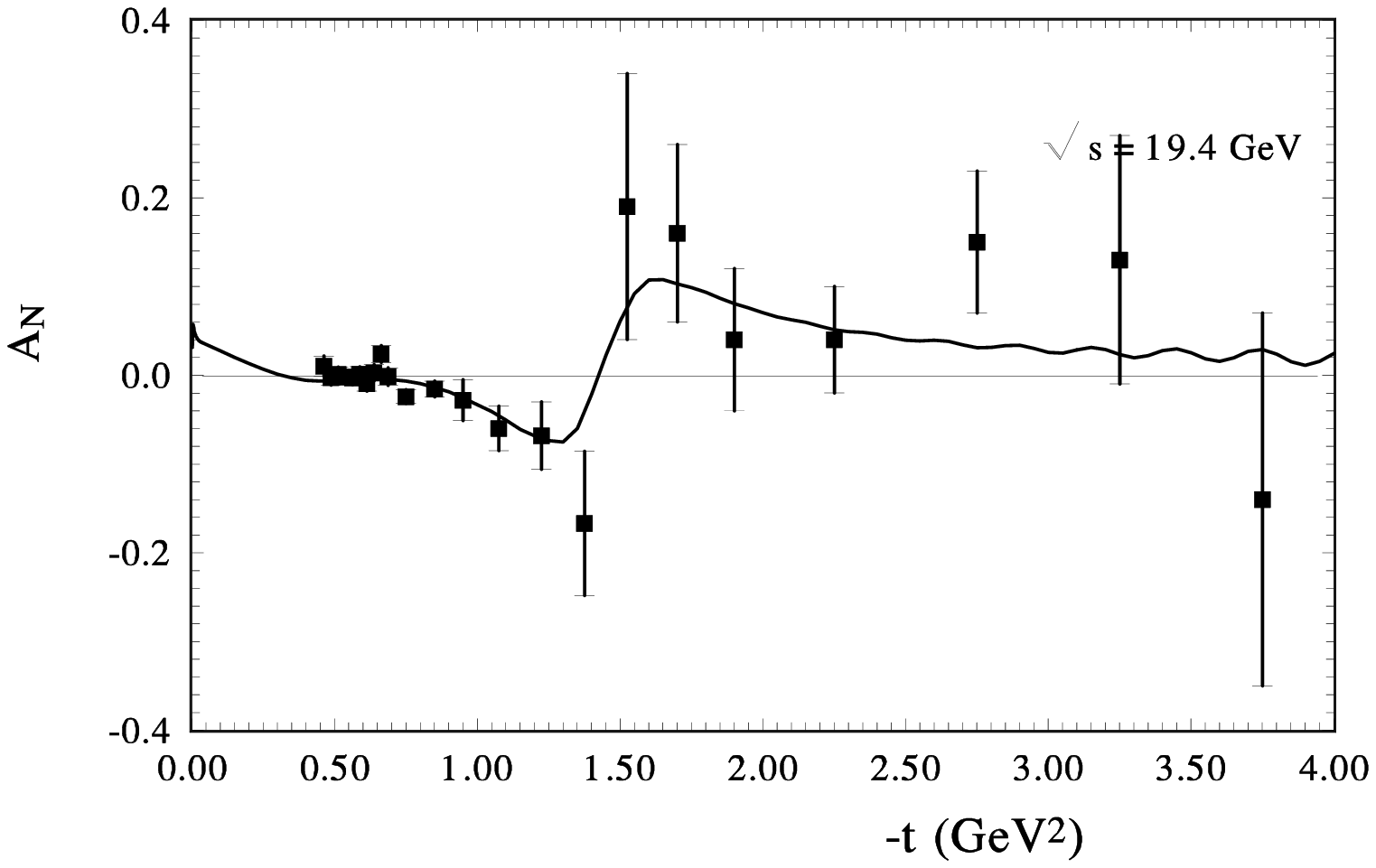}
\includegraphics[width=0.4\textwidth] {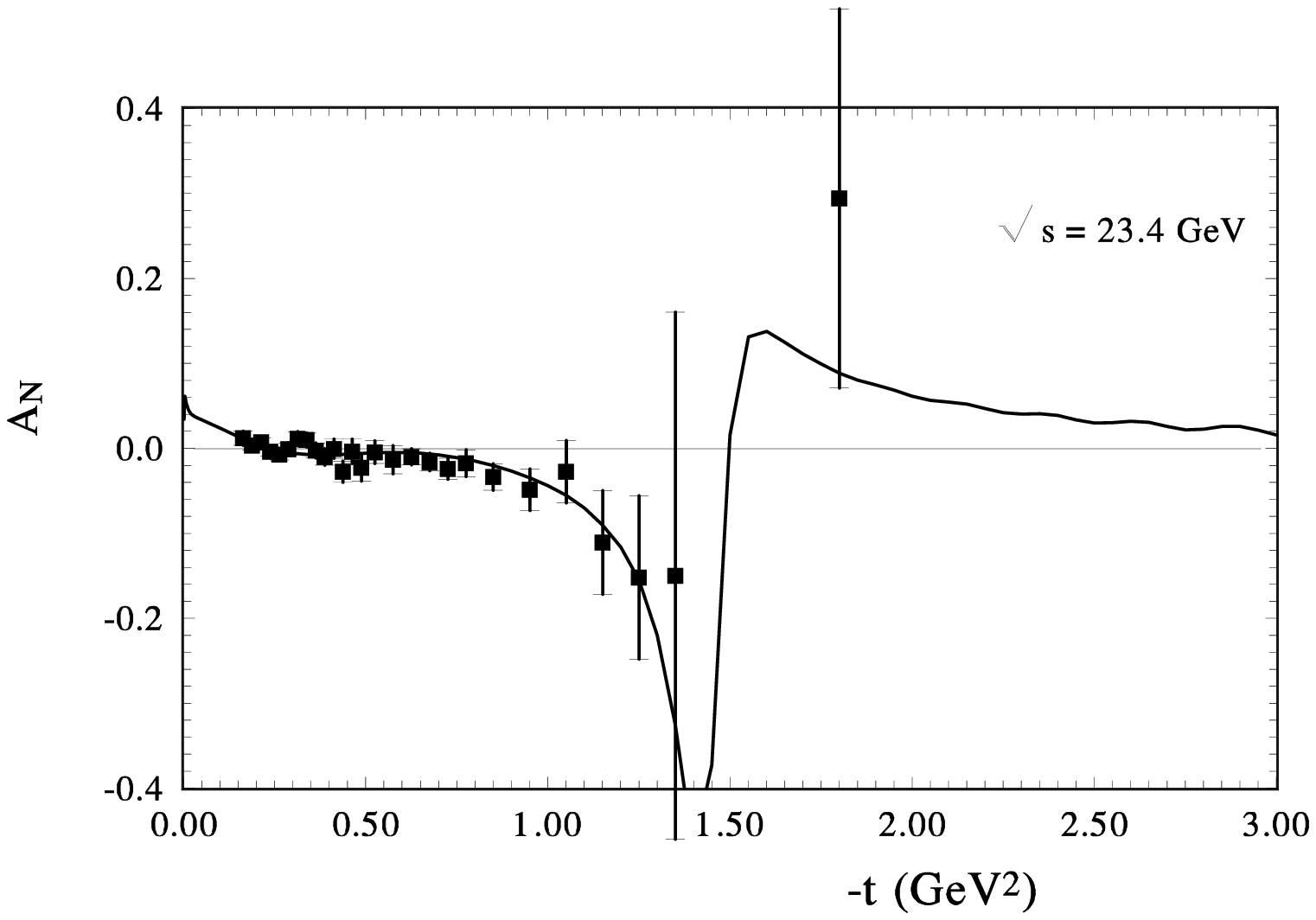}
\end{center}
\caption{The analyzing power $A_N$ of pp - scattering
      calculated:
   a) at $\sqrt{s} = 19.4 \ $GeV  (the experimental data \cite{Pol19p4}),
   and
   b)    at $\sqrt{s}= 23.4 \ $GeV
       (points - the existence experimental data \cite{Pol23p4})
          }
\label{fig:1}       % Give a unique label
\end{figure*}

\section{Conclusions}

     The Generelized parton distributions (GPDs) make it possible to better understand the thin hadron
     structure and  to obtain the hadron structure in the space frame (impact parameter representations).
     It is tightly connected with the hadron elastic hadron form factors.
    The research into the  form and energy dependence of the diffraction minimum of the differential cross sections
    of elastic hadron-hadron scattering at different energies
    will give  valuable information about the structure of the hadron scattering amplitude
     and  hence  the hadron structure and the  dynamics of strong interactions.
    The diffraction minimum corresponds to the change of the sign of the imaginary part of the spin-non-flip hadronic scattering amplitude
     and is created under a strong impact of the unitarization procedure.
    Its dip depends on the contributions of the real part of the spin-non-flip amplitude and the
    whole contribution of the spin-flip  scattering amplitude.
    In the framework of HEGS model, we show a deep connection between elastic and inealastic cross sections,
    which are tightly connected with the hadron structure at small and large distances.

    The HEGS model reproduces well the form and the energy dependence of the diffraction dip  of the proton-proton and proton antiproton
    elastic scattering \cite{Dif-min}. The predictions of the model in most part reproduce the form of the differential cross section at $\sqrt{s}=13 $ TeV.
    It means that the energy dependence of the scattering amplitude  determined in the HEGS model
     and unitarization procedure in the form of the standard eikonal
     representation satisfies the experimental data in the huge energy region (from $\sqrt{s}=9 $ GeV up to $\sqrt{s}=13 $ TeV.
     It should be noted that the real part of the scattering amplitude,
      on which the form and energy dependence of the diffraction dip heavily depend, is
     determined in the framework of the HEGS model only by the complex $\bar{s}$, and hence
      it is tightly connected with the imaginary part of the scattering amplitude
      and satisfies the analyticity and the dispersion relations.
%          Of course, the HEGS model is oversimplified, and to reproduce
            Quantitatively, for different thin structures of the scattering amplitude,
         a wider  analysis is needed.
          This concerns the fixed intercept taken from the deep inelastic processes and the fixed Regge slope $\alpha^{\prime}$,
           as well  as the form of the spin-flip amplitude.
           Such an analysis requires  a wider range of  experimental data, including the polarization data
        of $A_N(s,t)$, $A_{NN}(s,t)$, $A_{LL}(s,t)$, $A_{SL}(s,t)$.
%          and the normalized new data on the elastic $pp$ scattering at $\sqrt{s}=13$ TeV.
                        The obtained information about the sizes and energy dependence of the
                        spin-flip and double-flip amplitudes will make it possible
    to better understand % the behavior of the spin-depended differential cross section at large  angles.
     the results of  famous experiments carried out by A. Krish at the ZGS  to obtain the
          spin-dependent differential cross sections \cite{Krish1a,Krish1b} and
          the spin correlation parameter $A_{NN}$   %\cite{Krish1c}
          and   at the AGS \cite{Krish2} to obtain  the spin correlation parameter $A_{N}$
           showing the significant spin  effects at large momentum transfer.

%\vspace*{5mm}
%

\end{document}